\documentstyle [11pt] {article}
\makeatletter

\newcommand{\singlespacing}{\let\CS=\@currsize\renewcommand{\baselinestretch}{1}\tiny\CS}
\newcommand{\oneandahalfspacing}{\let\CS=\@currsize\renewcommand{\baselinestretch}{1.25}\tiny\CS}
\newcommand{\doublespacing}{\let\CS=\@currsize\renewcommand{\baselinestretch}{1.35}\tiny\CS}

\def\@citex[#1]#2{\if@filesw\immediate\write\@auxout{\string\citation{#2}}\fi
  \def\@citea{}\@cite{\@for\@citeb:=#2\do
    {\@citea\def\@citea{,\linebreak[0]\hskip0pt plus .2em}%
      \@ifundefined{b@\@citeb}%
      {{\bf ?}\@warning{Citation `\@citeb' on page \thepage\space undefined}}%
      \hbox{\csname b@\@citeb\endcsname}}}{#1}}

\newtheorem{rule-def}[theorem]{Rule}

\begin{document}

\newcommand{\la}{\lambda}
\newcommand{\si}{\sigma}
\newcommand{\ol}{1-\lambda}
\newcommand{\be}{\begin{equation}}
\newcommand{\ee}{\end{equation}}
\newcommand{\bea}{\begin{eqnarray}}
\newcommand{\eea}{\end{eqnarray}}
\newcommand{\nn}{\nonumber}
\newcommand{\lb}{\label}

\begin{center}

\large \bf{CLASSICAL ELECTRON MODEL WITH NEGATIVE ENERGY DENSITY IN
                            EINSTEIN-CARTAN THEORY OF GRAVITATION}
\end{center}

\begin{center}
 Saibal Ray$^\dagger$ \footnote{ E-mail: saibal@iucaa.ernet.in} and 
Sumana Bhadra$^*$\\
 $^\dagger${ Department of 
Physics, Barasat Government 
 College, Barasat 700 124, North 24 Parganas, West Bengal, India \&
 Inter-University Center for Astronomy and Astrophysics, 
Pune, India }\\
 $^*${Balichak Girls' High School, Balichak 721 124, West Midnapur,
 West Bengal, India} \\
 \end{center}

\vspace{0.25cm}
\noindent  
{\bf Abstract}\\
\noindent
Experimental result regarding the maximum limit of the radius of the
electron $\sim 10^{-16}$ cm and a few of the theoretical works suggest
that there might have some negative energy density regions within the 
particle in general theory of relativity. It is argued in the present 
investigation 
that such a negative energy density also can be obtained with a better 
physical interpretation in the framework of Einstein-Cartan theory.\\

\vspace{0.25cm}
\noindent
1.\quad INTRODUCTION\\
\noindent
Recently, Cooperstock and Rosen [1989], Bonnor and Cooperstock [1989], 
and 
Herrera and Varela [1994] have shown that within the experimentally 
obtained 
upper limit of the size of the electron ( $\sim 10^{-16}$ cm ) [Quigg, 
1983], 
when it is modeled as a charged sphere obeying Einstein-Maxwell theory, 
must 
contain some negative gravitational mass density regions within the 
particle . According to Cooperstock, Rosen and Bonnor (CRB) [1989], the 
rest mass or
active gravitational mass within this sphere, by virtue of the relation 
\be
M = m - \frac{q^2}{2a},
\ee
is negative and about $ 10^{-52} $ cm (when the inertial mass or 
effective
gravitational mass, charge and radius, respectively, of the electron, 
are $ m = 6.76 \times 10^{-56} $ cm, $ q = 1.38 \times 10^{-34} $ cm, 
and $ a = 10^{-16} $ cm in relativistic units ). Further, Herrera and
Varela (HV) [1994] have shown, in one of the cases of their paper, that 
the matter-energy density 
\be
\rho = ( \alpha q^2 + \frac{2}{3} \pi \sigma_{0}^2 )( a^2 - r^2 ),
\ee
for the constant $ \alpha = - 4.77 \times 10^{95} {\rm cm}^{-6} $ (when 
radius $ a
\sim 10^{-16} $ cm) is also negative, $\sigma_{0}$ being the constant 
charge
density at the centre of the spherical distribution. These models,
 however, lack spin and magnetic moment and hence do not possess the 
actual
physical characteristics required for an electron. \\
As an alternative way both the groups suggest the stationary 
Kerr-Newman (KN)
metric [Newman et al., 1965] related to the solution of
Einstein-Maxwell equations [Misner et al., 1973] to be more
appropriate than those described earlier. However, in this context it
is also to be mentioned here that the KN metric cannot be valid for 
distance scales of
the radius of a subatomic particle [Mann and Morris, 1993; Herrera and 
Varela, 1994].\\
We, therefore, feel that the problem can be tackled in the framework of
Einstein-Cartan (EC) theory, where torsion and spin are inherently 
present in
the formulation of the theory itself.\\

\noindent
2.\quad AN OVERVIEW: THE NEGATIVE DENSITY MODELS\\
\noindent
Before going into the Einstein-Cartan theory let us have a birds' eye
view of the negative matter-energy density models already 
we have mentioned in the introduction.\\
\noindent
(i)\quad THE COOPERSTOCK-ROSEN-BONNOR (CRB) MODEL\\
\noindent
Cooperstock and Rosen [1989] and Bonnor and Cooperstock [1989] in
their papers have shown that any spherically
symmetric distribution of charged fluid, irrespective of its equation
of state, whose total mass, radius and charge correspond to the
observed values of the electron, must have a negative energy
distribution (at least for some values of the radial coordinate).  
Considering a static spherically symmetric charge distribution with
the line-element 
\be
ds^{2}= e^{2\nu(r)}dt^{2} - e^{2 \lambda (r)} dr^{2} - r^{2}(d 
\theta^{2}+sin^{2} \theta d \phi^{2})
\ee
they have argued that when the Einstein-Maxwell equation  
\be
{R^{0}}_0 - \frac{1}{2}{\delta^{0}}_0 R = 8\pi ({{T^{0}}_0}^{(m)} +
{{T^{0}}_0}^{(em)})
\ee
is written in the form
\be
e^{-2\lambda} = e^ {2\nu} = 1 - \frac{1}{r} \int_{0}^{r}( 8\pi\rho +
e^{-(\nu + \lambda)} E^2) r^2 dr
\ee 
and hence is equated with the Reissner-Nordstr\"{o}m exterior
metric on the boundary $r = a$, which as usual gives
\be
 1 - \frac{2m}{a} + \frac{q^2}{a^2} = 1 - \frac{1}{a} \int_{0}^{a}( 
8\pi\rho +
e^{-(\nu + \lambda)} E^2) r^2 dr .
\ee
Then for the previous specifications of mass, charge and radius of the
electron it can be shown that 
\be
\frac{q^2}{a^2} - \frac{2m}{a} \sim 2\times 10^{-36} > 0.
\ee
So, the left hand side of the above equation (6) must be greater than
unity and hence on the right hand side $\rho < 0 $ for some values of
$r$ implying that the electron must contain some negative rest mass
density though the net mass is as usual a positive quantity. \\
\noindent
(ii)\quad THE HERRERA-VARELA (HV) MODEL\\
\noindent
Following the CRB model [1989] Herrera and Varela [1994] have
discussed the fact that the electron, when modeled as
a relativistic spherically symmetric charged distribution of matter,
must contain
some negative rest mass if its radius is not larger than $\sim 
10^{-16}$
cm. In this regard they have analyzed some extended electron models
and have shown that negative energy density distributions result from
the requirement that the total mass of these models remains constant
in the limit of a point particle. Among all these extended electron
models the model of Tiwari et al. [1984] demands special attention to 
us which
will be seen very much relevant to our present work. Herrera and
Varela [1994] generalize this model of Tiwari et al. [1984] by
introducing a condition of anisotropy in the form
\be
p_{\perp} - p_r = \alpha q^2 r^2
\ee
 where  $\alpha$ is a constant. \\
\noindent
Thus the solution obtained by Herrera and Varela [1994] is as follows:
\be
e^{-2\lambda} = e^ {2\nu} = 1 - \frac{16}{45}{\pi}^2 {\sigma_{0}}^2 r^2
( 5 a^2 - 2 r^2 ) - \frac{8}{15}\pi \alpha q^2 r^2 (5 a^2 - 3 r^2 ) ,\\ 
\ee
\be
p_r = - ( \alpha q^2 + \frac{2}{3}\pi {\sigma_{0}}^2) ( a^2 - r^2 ) ,\\
\ee
\be
p_{\perp} = \alpha q^2 r^2 - (\alpha q^2 + \frac{2}{3}\pi
{\sigma_{0}}^2) ( a^2 -	 r^2) ,\\
\ee
\be
m = \frac{64}{45} {\pi}^2 {\sigma_{0}}^2 a^5 + \frac{8}{15}\pi \alpha
q^2 a^5 ,\\
\ee
\be
\rho = (\alpha q^2 + \frac{2}{3}\pi {\sigma_{0}}^2) ( a^2 - r^2 ).\\
\ee
\be
q = \frac{4}{3}\pi \sigma_{0} a^3  
\ee
The value of $\alpha$ can be obtained from the equation (12) as
$\alpha = - 4.77 \times 10^{95} {\rm cm}^{-6}$ and
hence the energy density, as given by the equation (13) is negative for
the radius of the electron $a = 10^{-16} $cm.\\
\noindent
Now, from the equation (12) it can be seen that the effective
gravitational mass, $m$, is of purely
electromagnetic origin and corresponds to the TRK model [1984] with
$\alpha = 0 $ case. This type of models where mass, including all the
other physical parameters, originates from the
electromagnetic field alone are known as the electromagnetic mass 
models [EMMM]
in the literature [Feynman et al., 1964] and have been investigated by
several  authors [Florides, 1962, 1983; Cooperstock and de la Cruz,
1978; Tiwari et al., 1984, 1986, 1991, 2000; Gautreau, 1985; Gr{\o}n, 
1985,
1986a, 1986b; de Leon, 1987a, 1987b, 1988; Tiwari and Ray, 1991a,
1991b, 1997; Ray et al., 1993; Ray and Ray, 1993]. In the present
paper we shall construct such a model within the framework of
Einstein-Cartan theory with negative matter-energy density for some 
values of the radial coordinate.\\

\noindent
3.\quad THE FIELD EQUATIONS OF EINSTEIN-CARTAN THEORY \\
\noindent
The EC field equations are given by   
\be
 {R^{i}}_{j} - \frac{1}{2}{\delta^{i}}_{j} R = -\kappa {t^{i}}_{j} ,
\ee
\be
{Q^{i}}_{jk} - {\delta^{i}}_{j} {Q^{l}}_{lk} - 
{\delta^{i}}_{k}{Q^{l}}_{jl} = - \kappa {S^{i}}_{jk},
\ee
where $ {t^{i}}_{j} $ is the canonical energy-momentum tensor 
(asymmetric),
$ {Q^{i}}_{jk} $ is the torsion tensor and $ {S^{i}}_{jk} $ is the spin
tensor (with $ \kappa = - 8 \pi $, $ G $ and $ c $ being unity in 
relativistic
units ).\\
The asymmetric energy-momentum tensor here is given by
\be
{t^{i}}_{j} = {T^{i}}_{j} + \frac{1}{2} g^{ik} \nabla_{m} 
({S^{m}}_{jk}),
\ee
$ \nabla_{m}$  being covariant derivative with respect to the
torsionless, symmetric Levi-Civita  connection $ {\Gamma^{i}}_{jk}$ and 
the 
symmetric energy-momentum tensor ${ T^i}_j$ will consist of two parts, 
viz.,
matter and electromagnetic tensors and which, respectively, are
\be
{{T^{i}}_{j}}^{(m)} = (\rho + p)u^{i} u_{j} - p{g^{i}}_{j},\\
\ee
\be
{{T^{i}}_{j}}^{(em)} = \frac{1}{4 \pi} (- F_{jk} F^{ik} + \frac{1}{4}
{\delta ^{i}}_{j} F_{kl} F^{kl} ) ,
\ee
where $ \rho $ is the matter-energy density, $p$ is the fluid
pressure, $ u^{i}$ is the velocity 
four-vector (with $ u^{i} u_{i} = 1$) and $ F_{ij} $ is the
electromagnetic field tensor.\\
\noindent
The conservation equations for the EC theory can be given through the
Bianchi identities as 
\be
\nabla_{k}[(\rho + p) u^{k} - g^{ki}u^{l} \nabla_{m}(u^{m} S_{li})] = 
u^{j}
\nabla_{j}p, 
\ee
\be
[(\rho + p)u^{k} - g^{ki}u^{l} \nabla_{m}(u^{m}S^{li})] \nabla_{k}u_{j}
= - \nabla_{l}(u^{l}u_{j}) + u^{k}S_{jm}{R^{m}}_{k} - \frac{1}{2}
u^{k}S_{lm}{R^{lm}}_{jk} 
\ee
\noindent
Now, electromagnetic fields not being coupled with torsion [Novello,
1976; Raychaudhuri, 1979]
the Maxwell equations as usual take the form
\be
\nabla _{j} F^{ij} = J^{i},\\
\ee
\be
( J^{i} \sqrt{- g} )_{,i} = 0.
\ee
The electromagnetic field tensor, $ F_{ij}$, in the above equation (22) 
is 
related to the electromagnetic potentials as $ F_{ij} = A_{i,j} - 
A_{j,i} $ which
is equivalent to $ F_{[i,j,k]} = 0 $, $A_{i}$ being the electrostatic
potentials. 
Here and in what follows a comma denotes the partial derivative with 
respect
to the coordinate indices involving the index.\\
Again, the spin tensor and the intrinsic angular momentum density 
tensor are
related as 
\be
{S^{i}}_{jk} = u^{i}S_{jk},
\ee
with
\be 
u^{i}S_{ik} = 0.
\ee
Now, assuming that the spins of the individual charged particles 
composing the
fluid distribution are all aligned in the radial directions [Prasanna,
1975; Raychaudhuri, 1979; Tiwari and Ray, 1997] and the matter is at
rest with 
respect to the observer, the non-vanishing components of 
the spin tensor can be obtained, from equations (24) and (25), as 
\be
{S^{0}}_{23} = - {S^{0}}_{32} = s(g_{00})^{- 1/2},
\ee
whereas, from equation (16), we have the torsion tensor as 
\be
{Q^{0}}_{23} = - {Q^{0}}_{32} = - {\kappa} s(g_{00})^{- 1/2},
\ee
$s = S_{23}$ being the only non vanishing component of the intrinsic 
angular
momentum density tensor. Here, we have followed the convention $(t, 
r,\theta,
\phi) = (0, 1, 2, 3)$.\\
\noindent
The Einstein-Cartan-Maxwell equations with source then can be written
as [Tiwari and Ray, 1997]
\be
e^{-2 \lambda} (\frac{2 \lambda^\prime }{r} - \frac{1}{r^2}) + 
\frac{1}{r^2} = 8 \pi \tilde{\rho} + E^2,
\ee
\be
e^{-2 \lambda} (\frac{2 \nu^\prime}{r} + \frac{1}{r^2}) - \frac{1}{r^2}  
= 8 \pi \tilde{p_r} - E^2,
\ee
\be
e^{- 2\lambda} [\nu^{\prime\prime} + {\nu^{\prime}}^2
-\nu^{\prime}\lambda^{\prime}  + \frac{( \nu^{\prime} 
-\lambda^{\prime})}{r}] =
8 \pi \tilde{p_\perp} + E^2,
\ee
\be
(r^2 E)^{\prime} = 4 \pi r^2 \sigma e^{\lambda},
\ee
where \quad $ \tilde{\rho},  \quad \tilde{p_r}, \quad \tilde{p_\perp}$ 
and \quad$ E$ are the
effective matter-energy density, effective pressures (radial and 
tangential)
and electric field respectively, and are defined as \\
\be
\tilde {\rho} = \rho -  2 \pi s^2, 
\ee
\be
\tilde {p}_r = p_r - 2 \pi s^2,
\ee
\be
\tilde {p}_{\perp} = p_{ \perp} - 2 \pi s^2,
\ee
\be
E = - exp[ -(\nu + \lambda)] {\phi}^\prime = \frac{q}{r^2},
\ee
$ \rho, \quad  p_r,  \quad p_{\perp},  \quad s, \quad  \phi $ and \quad 
$ q $ being the ordinary matter-energy
density, ordinary pressures (radial and tangential), spin density,
electrostatic potential and electric charge respectively. Here, 
$\sigma$
denotes the electric charge density and prime is used for the 
derivative with
respect to the radial coordinate $r$.\\ 
\noindent
Also, the conservation equations (20) and (21) here reduce to 
\be
\frac{d \tilde{p}_r}{dr} = -(\tilde {\rho} + \tilde {p}_r) \nu^{\prime} 
+ \frac{1}{8 \pi r^4} \frac{d}{dr} (q^2) +\frac{ 2(\tilde {p}_{\perp} -
\tilde {p}_r)}{r}.
\ee
This is the key equation which is to be solved for constructing EMM.\\

\noindent
4.\quad THE SOLUTIONS \\
\noindent
Addition of (28) and (29), under the assumption $ g_{{0}{0}} g_{{1}{1}}
 = - 1 $ ( or equivalently, in terms of energy-momentum tensors 
${T^0}_0 =
{T^1}_1 $), provides the pure charge condition 
\be
\tilde{\rho} + \tilde{p}_r = 0,
\ee
where, in general, $\tilde{\rho} $ is assumed to be positive and hence 
$ \tilde
{p}_r $ is negative. However, as is evident from equation (32), $
\tilde{\rho}$, being the effective energy-density, can even be negative 
due to the
positive second term related to the spin on the right hand side. Thus, 
the
possibility of equation (33) being satisfied with $\tilde{p_r}$ being 
positive
is not ruled out. \\
\noindent
To make (31) and (36) solvable, we further assume that 
\be
\sigma e^{\lambda}  = \sigma_{0},\\
\ee
\be
p_{\perp} - p_r = \tilde {p}_{\perp} - \tilde{p}_r = \alpha q^2 r^2
\ee
following TRK Model [Tiwari et al., 1984] and HV model [Herrera and
Varela, 1994] respectively, where 
$\sigma_{0}$ and $ \alpha$ are two constants as mentioned earlier. \\
\noindent
By substituting (37) -- (39) in (31) and (36), we get 
\be
E = \frac{q}{r^2}=\frac{4}{3}\pi \sigma_{0} r  ,\\
\ee
\be
p_r = 2 \pi s^2 - ( \alpha q^2 + \frac{2}{3}\pi {\sigma_{0}}^2) ( a^2
- r^2 ),\\
\ee
\be
p_{\perp} = 2 \pi s^2 - \alpha q^2 ( a^2 - 2 r^2 ) - \frac{2}{3}\pi
{\sigma_{0}}^2 ( a^2 - r^2)  ,\\
\ee
\be
\rho = 2 \pi s^2 + (\alpha q^2 +\frac{2}{3}\pi {\sigma_{0}}^2) ( a^2 - 
r^2 ).
\ee
\noindent 
The active gravitational mass 
\be
M(r) = 4 \pi \int_{0}^{r}( \rho - 2 \pi s^2 + \frac{E^2} {8 \pi} ) r^2 
dr
\ee
takes the form, by virtue of (40) and (43), as 
\be
M(r) = \frac{8}{135}{\pi}^2 {\sigma_{0}}^2 r^3 [ 8 \pi \alpha a^6 ( 5 
a^2 - 3r^2 ) + 3(
5a^2 - 2r^2 ) ].     
\ee
\noindent
Thus, the metric potentials $\lambda$ and $ \nu$ are given by 
\be
e^{-2\lambda} = e^ {2\nu} = 1 - \frac{2M(r)}{r}, 
\ee
whereas, the effective gravitational mass in (1), can be obtained as 
\be
m =\frac{64}{45}  {\pi}^2 {\sigma_{0}}^2 a^5 ( 1 + \frac{2}{3}\pi 
\alpha a^6   ),
\ee
which corresponds to the second case (B) of HV model [1994] and of 
purely
electromagnetic origin. This again corresponds to the TRK model [1984] 
with
$\alpha = 0 $ case. It can be noted here that unlike the matter-energy
density the effective gravitational mass is independent of spin.\\ 
\noindent
In this context it is to be mentioned here that the junction
conditions in 
the EC theory are different from that of general 
theory of relativity and indeed read like this [Arkuszewski et al., 
1975]
\be
n_{i} u^{i} \mid _{-} = 0\\
\ee
\be
p \mid _{-} = 2 \pi G ( n_{i} S^{i} ) \mid _{-}
\ee
where $ S^{i}$ is the spin density pseudo-vector. Here condition (48) 
is the
same as in classical relativistic hydrodynamics and has already  been
incorporated by matching the interior solution to the exterior
Reissner-Nordstr\"{o}m field at the boundary of the spherical 
distribution.
The condition (49), however, is the additional condition to be 
satisfied in EC
theory. In the present case, it is only the effective pressure (radial) 
that 
vanishes on the boundary and not the ordinary radial pressure which, by 
virtue
of equation (41), equals  $ 2 \pi s^{2} $. The spin is aligned in the 
radial 
direction and hence the spin density pseudo-vector is hypersurface 
orthogonal. 
Thus, the boundary condition (49) will become
\be
p_{r}{\mid}_{r = a - 0} = 2 \pi s^2 {\mid}_{r = a - 0},
\ee
which, depending on whether $s$ is a constant or function of 
coordinates, will
automatically be satisfied.\\
\noindent
In this connection it is to be mentioned here that the spin 
density,$s$, in the final 
solutions (41) -- (43) remains arbitrary (function of $r$). An explicit 
functional form of this spin density can also be obtained by assuming 
some additional
 physically viable possibility, such as the one used by Prasanna [1975] 
by 
splitting the conservation equation into two parts, the second part
relating to conservation 
of spin only, giving the functional form of spin density as $ s = s_{0}
 e^{-\nu}$ (where $s_{0}$ is the value of $s$ at $r = 0$ i.e. the
central spin density). This can, using equations
(38) and (46) and the condition $g_{oo}g_{11} = 1$, (that is, $ \nu + 
\lambda = 0 $)
 equivalently be written as $ s \sigma = s_{0} \sigma_{0} =$ constant. 
The 
functional form of spin density is, however, not relevant in our 
discussion as our 
problem is concerned with the properties related to the `electron', an 
elementary particle whose radius is of the order of $ 10^{- 16}$ cm. 
Indeed,
 as the spin function is arbitrary, there is no loss in generality, 
even if we 
assume it to be almost a constant (that is, the quantized value of the 
spin of
 the electron).\\

\noindent
6.\quad THE NEGATIVE ENERGY DENSITY MODEL\\
\noindent 
Let us have a closer observation of the results of the previous section
$4$. The equation (43) related to matter-energy density has the spin
density part in the first term where spin density is defined as $s =
3S/4\pi a^{3}$, where $S$ is the spin of electron the quantized value
of which is $S = \hbar/2$. Then substituting the standard values for 
different 
parameters in the relativistic units, as mentioned in the introductory
part, the numerical value for the matter-energy density (43) can be 
shown as 
\be
\rho = 6.14\times 10^{-37} - 6.81\times 10^{27} ( 10^{-32} - r^2 ).
\ee
\noindent
The first term related to spin, being of the order of $ 10^{-37}$,
is too small compared to the last term and hence the spin contribution
is negligible. Now, the equation (51) indicates that the central 
density at
$ r = 0 $ is negative and its magnitude is about $10^{-5}$. On the
other hand, the total density at the boundary, $ r = a$,
is positive as usual with the numerical value about $10^{-37}$. This
change in the sign of the energy density is because of the presence of 
the spin
term in equation (52) which, indeed, is the contribution of the EC
theory. In the absence of spin, however, we could have negative and
zero densities at the centre and boundary of the electron
respectively. \\ 
\noindent
This change in the sign again indicates that the central negative
value gradually increases along the radius and somewhere, in the
region $0<r_{c}<a$, it becomes zero, where $r_{c}$ is the critical
radius. Obviously, the amount of negative energy density is less than
its positive counterpart the balance of which ultimately provides the
net density as the positive one.\\
\noindent
It is already mentioned that though, in general, for any spherical
fluid distribution the density on the surface should be zero we are
getting here some non-zero value for it. This finite value is solely
coming from the spin contributed part $2\pi s^{2}$. Thus for the
vanishing spin the situation corresponds to the general behaviour
(vide equation (17) of Herrera and Varela, 1994). In this context it
is also to be noted here that up to the critical radius behaviour of 
our model is
similar to that of Herrera and Varela [1994]. Beyond this cut off
radius the energy density is regulated by spin which makes the overall
density of the model as positive. This particular aspect lack in the
model of Herrera and Varela [1994] where the total energy density is a
negative quantity. This increase of
matter-energy density due to spin density can probably be accounted
for the kinetic energy through the angular motion of the electron
here.\\
Similar kind of examination is also possible for the pressures, radial
and tangential, both. The radial pressure, in this case, takes the
form as    
\be
p = 6.14\times 10^{-37} + 6.81\times 10^{27} ( 10^{-32} - r^2 ).
\ee
However, pressure is throughout positive here from the centre to
boundary. These results, i.e. negative energy density and positive
pressure, are in accordance with the pure charge condition (37) which
reads as $\rho= - p_r + 4 \pi s^2$ .\\

\noindent
7.\quad CONCLUSIONS\\
\noindent
(i) The possible origin of the intriguing negative matter-energy
density in the work of Cooperstock and Rosen [1989], Bonnor and
Cooperstock [1989], Herrera and Varela [1994] and present paper may be 
due to 
 the finiteness of the total mass of the Reissner-Nordstr\"{o}m
solution [Papapetrou, 1974; Visser, 1989]. Since the electrostatic
energy of a point charge is infinite, the only way to produce a finite
total mass is the presence
of an infinite amount of negative energy at the center of symmetry.
According to Bonnor and Cooperstock [1989] the negativity of the
energy density and hence the active gravitational mass is consistent
with the phenomenon, known as the Reissner-Nordstr\"{o}m repulsion
[de la Cruz and Israel, 1967; Cohen and
Gautreau, 1979; Tiwari et al., 1984; Cooperstock and Rosen, 1989].   
In this regards Bonnor and Cooperstock [1989] also have discussed about 
the
singularity theorems of general relativity [Hawking and Ellis,
1973]. They have shown that the negative regions are liable to exit
over distances of order $10^{-13}$ cm and as the proof of the
singularity theorems depends on the manifold structure of
space-time valid down to lengths of order $10^{-15}$ cm so might break
down below this. On the other hand, in the context of Einstein-Cartan
theory of gravitation the idea of negative mass is not a new one as 
stated by Sabbata and Sivaram:``...torsion
provides a natural framework for the description the negative mass
under extreme conditions of such as in the early universe, when a
transition from positive to negative mass can take place.''\\    
\noindent
(ii) We have considered in the present paper an extended static 
spherically 
symmetric distribution of an elementary particle like electron having 
the 
radius of the order of $ 10^{- 16}$ cm, and even if for the finite size 
of the 
physical system the spin in Einstein-Cartan theory can be related to 
orbital
rotation (which indeed is not the case), for systems of dimensions of 
subatomic particle the orbital rotation loses its meaning. In this 
case, the 
only way is to take the spin to be the `intrinsic angular momentum', 
that is, 
the spin of quantum mechanical origin (in our problem since $ s $ is 
arbitrary,
we can consider its quantized value or an average value). In this
respect we would 
like to quote here from Hehl et al. [1974],``It is crucial to note
that
 {\it{spin}} in $U_{4}$ theory is canonical spin, that is, the
{\it{intrinsic}} 
spin of elementary particles,
 not the so-called spin of galaxies or planets."\\
\noindent
(iii) Following other authors [Prasanna, 1975; Raychaudhuri, 1979;
Tiwari and Ray, 1997], in the present work the spins of all 
the individual particles are assumed to be oriented along the radial 
axis of 
the spherical systems. As to how this alignment is brought about is not 
very 
much clear. We have discussed here only a few possible ways of 
realizing this
situation. According to Raychaudhuri [1979], in general, there will be 
an
interaction between the spins of the particles and the magnetic field. 
The
overall effect is the alignment of the spins. In this context, Prasanna 
[1975]
mentions that such an alignment may be meaningful either in the case of
spherical symmetry when magnetic field is present or else one has to 
consider
axially symmetric field. As stated above, our view point is that in the 
case of
physical systems of the size of the electron, the radial alignment of 
the spin 
is not ruled out. The solution obtained supports this view.\\
\noindent
(iv) Though our present approach via Einstein-Cartan theory to inject 
spin may be interesting, we feel even that there should have some room 
to discuss the relationship of our work with an alternative means to 
provide spin and magnetic moment. This is, we think, possible through 
Dirac-Maxwell theory where spin and magnetic moment are naturally 
incorporated through the Dirac spin [Bohun and Cooperstock, 1999; Lisi, 1995] 
and would like to persue the problem in future investigations.

\vspace{0.5cm}

\noindent
Acknowledgment\\
\noindent
One of the authors (SR) thanks are due to the authority of IUCAA, 
Pune, India for providing Associateship Programme under which a part of this 
work was carried out. The authors are also grateful to the  referee  for 
the critical comments which made it possible to improve the paper. \\
\vspace{0.25cm}

\noindent
{\bf {References}}\\
\noindent
Arkuszewski, W., Kopczynski, W. and Ponomariev,
              V. N. (1975). {\it{Comm. Math.  Phys.}} {\bf{45}},
              183.\\
Bohun, C. S. and Cooperstock, F. I. (1999). {\it{Phys. Rev. A}} 
{\bf{60}}, 4291.\\
Bonnor, W. B. and Cooperstock, F. I. (1989). {\it{Phys. Lett. A}}
              {\bf{139}}, 442.\\
Cohen, J. M. and Gautreau, R. (1979). {\it{Phys. Rev. D}} {\bf{19}}, 
2273.\\
Cooperstock, F. I. and de la Cruz, V. (1978). {\it{Gen. Rel. Grav.}}
              {\bf{9}}, 835.\\
Cooperstock, F. I. and Rosen, N. (1989). {\it{Int. J. Theo. Phys.}}
              {\bf{28}}, 423.\\
de la Cruz, V. and Israel, W. (1967). {\it{Nuo. Cim.}} {\bf{51}},
              744.\\
de Sabbata, V. and Sivaram, C. (1994). {\it{Spin and Torsion in
              Gravitation}} (World Scientific, Singapore, Chap. VII).\\
Feynman, R. P., Leighton, R. R. and Sands, M. (1964). {\it{The Feynman
              Lectures on Physics}} (Addison-Wesley, Palo Alto,
              Vol.II, Chap. 28). \\
Florides, P. S. (1962). {\it{Proc. Camb. Phil. Soc.}} {\bf{58}}, 102; 
(1983). {\it{Phys. A: Math. Gen.}} {\bf{16}}, 1419.\\
Gautreau, R. (1985). {\it{Phys. Rev. D}} {\bf{31}}, 1860.\\
Gr{\o}n, {\O}. (1985). {\it{Phys. Rev. D}} {\bf{31}}, 2129; (1986a).
		{\it{Am. J. Phys.}} {\bf{54}}, 46; (1986b). {\it{Gen. Rel. Grav.}} 
{\bf{18}}, 591.\\	
Hawking, S. W. and Ellis, G. F. R. (1973). {\it{The Large Scale
              Structure of Space-time}} (Cambridge University Press).\\
Hehl, F. W., von der Heyde, P. and Kerlick, G. D. (1974). {\it{Phys. 
Rev. D}} {\bf{10}}, 1066.\\
Herrera, L and Varela, V. (1996). {\it {Phys. Lett. A}} {\bf {189}},11. 
\\
Lisi, E (1995). {\it{Journal of Physics A}} {\bf{28}}, 5385.\\
Mann, R. and Morris, M. (1993). {\it{Phys. Lett. A}} {\bf{ 181}},
              443. \\
Newman, E., Couch, F., Chinapared, K., Exton, A., Prakash, A. and 
Torrence,
          R. (1965). {\it{J. Math. Phys.}} {\bf{6}}, 918.\\
Novello, M. (1976). {\it{Phys. Lett. A}} {\bf{59}}, 105.\\
Ponce de Leon, J. (1987a). {\it{J. Math. Phys.}} {\bf{28}}, 410; 
(1987b). 
            {\it{Gen. Rel. Grav.}} {\bf{19}}, 797;
           (1988). {\it{J. Maths. Phys.}} {\bf{29}}, 197.\\
Prasanna, A. R. (1975). {\it{Phys. Rev. D }} {\bf{11}}, 2076.\\
Quigg, C. (1983). {\it{Gauge Theories of the Strong, Weak and 
              Electromagnetic Interactions}} (Benjamin, N.Y.), P. 3.\\
Ray, S., Ray, D. and Tiwari, R. N. (1993). {\it{Astrophys. Space
          Sci.}} {\bf{199}}, 333.\\
Ray, S. and Ray, D. (1993). {\it{Astrophys. Space Sci.}} {\bf{203}},
            211.\\
Raychaudhuri, A. K. (1979). {\it{Theoretical Cosmology}} (Clarendon
              Press, Oxford) Chap. 10 .\\
Tiwari, R. N., Rao, J. R. and Kanakamedala, R. R. (1984). {\it{Phys.
		Rev. D}} {\bf{30}}, 489.\\
Tiwari, R. N., Rao, J. R. and Kanakamedala, R. R. (1986). {\it{Phys.
		Rev. D}} {\bf{34}}, 1205.\\
Tiwari, R. N., Rao, J. R. and Ray, S. (1991). {\it{Astrophys. Space
		Sci.}} {\bf{178}}, 119.\\
Tiwari, R. N. and Ray, S. (1991a). {\it{Astrophys. Space Sci.}} 
{\bf{180}}, 
		143;  (1991b). {\it{Astrophys. Space Sci.}} {\bf{182}}, 105.\\
Tiwari, R. N. and Ray, S. (1996). {\it{Ind. J. Pure Appl. Math.}} 
{\bf{27}}, 907.\\
Tiwari, R. N. and Ray, S. (1997). {\it{Gen. Rel. Grav.}} {\bf{29}}, 
683.\\
Tiwari, R. N., Ray, S. and Bhadra, S. (2000). {\it{Ind. J. Pure Appl. 
Math.}}               {\bf{31}}, 1017.\\ 
Visser, M. (1989). {\it{Phys. Lett. A}} {\bf{139}}, 99.\\

\end{document}